\documentclass[prd,10pt,aps,showpacs,twocolumn,superscriptaddress,nofootinbib,floatfix,amssymb,amsmath]{revtex4-1}
\usepackage{slashed}
\usepackage{epsfig}

\newcommand{\beqn}{\begin{eqnarray}}
\newcommand{\eeqn}{\end{eqnarray}}

\newcommand{\cE}{{\cal E}}

\newcommand\abs[1]{\left|#1\right|}

\usepackage{color}

\newcommand{\revision}[1]{{#1}}

\begin{document}

\title{Schwinger pair creation in Dirac semimetals in the presence of external magnetic and electric fields}

\author{R.A. Abramchuk}
\affiliation{Moscow Institute of Physics and Technology, 9, Institutskii per., Dolgoprudny, Moscow Region, 141700, Russia}

\author{M.A. Zubkov\footnote{on leave of absence from Moscow Institute of Physics and Technology, 9, Institutskii per., Dolgoprudny, Moscow Region, 141700, Russia}}
\email{zubkov@itep.ru}
\affiliation{LE STUDIUM, Loire Valley Institute for Advanced Studies,
Tours and Orleans, 45000 Orleans France}
\affiliation{Laboratoire de Mathematiques et de Physique
Theorique, Universite de Tours, 37200 Tours, France}
\affiliation{Institute for Theoretical and Experimental Physics, B. Cheremushkinskaya 25, Moscow, 117259, Russia}
\affiliation{National Research Nuclear University MEPhI (Moscow Engineering
Physics Institute), Kashirskoe highway 31, 115409 Moscow, Russia}

\begin{abstract}
We discuss the Schwinger pair creation process for the system of massless Dirac fermions in the presence of constant external magnetic and electric fields. The pair production rate remains finite unlike the vacuum decay rate. In the recently discovered Dirac semimetals, where the massless Dirac fermions emerge, this pair production may be observed experimentally through the transport properties. We estimate its contribution to the ordinary conductivity of the semimetals.
\end{abstract}

\pacs{}

\date{\today}

\maketitle

\section{Introduction}

{Recently the Dirac \cite{semimetal_discovery,semimetal_discovery2,semimetal_discovery3,ZrTe5,ZrTe5:2,Bi2Se3} and Weyl \cite{WeylSemimetalDiscovery} semimetals were discovered experimentally. This discovery is important for the development of the inter - relation between the condensed matter physics and the high energy physics because those materials (as well as $^3$He-A \cite{Volovik2003}) are able to serve as an arena for the experimental investigation of various effects specific for the high energy physics \cite{semimetal_effects10,semimetal_effects11,semimetal_effects12,semimetal_effects13,ref:semimetal:4,Zyuzin:2012tv,tewary}.}
{The low energy effective theory of fermionic quasiparticles in Dirac and Weyl semimetals has an emergent relativistic invariance \cite{Volovik2003,Chernodub:2015wxa}. Fermions in Dirac and Weyl semimetals are charged, and they are influenced by the external magnetic and electric fields. This extends the possibility to observe various effects related to chiral anomaly \cite{ref:diffusion,ref:transport,semimetal_discovery3,ref:semimetal:4}.}

{The ordinary chiral anomaly is related to the spectral flow of the one - particle Hamiltonian along the Lowest Landau Level (LLL). This effect dominates if $E/B \ll 1$. The Schwinger pair creation process related to the remaining Landau levels represents the next approximation of the expansion in this parameter. In the present paper we consider this effect and estimate its contribution to the total conductivity\footnote{Actually, the anomalous pumping of quasiparticles from vacuum at the lowest Landau level may be considered as the degenerate case of the pair production process, that occurs with probability equal to $1$.}.}
{It is worth mentioning, that the rate of the decay of vacuum and the rate of the Schwinger pair creation process \cite{SMR} are different quantities. Vacuum persistence probability is equal to the exponent of the vacuum decay rate (with the minus sign). The pair production rate contributes to this quantity only partly. The vacuum decay rate (which is equal to the imaginary part of effective action) was calculated in a number of ways for the Dirac particles with finite mass $m$ (see, for example, \cite{Schwinger,Sauter}). The corresponding \revision{pair} production rate in the presence of constant electric and magnetic fields was calculated in \cite{Nikishov} using the exact solution of Dirac equation.  In \cite{SPS} the same result was reproduced using the semi - classical technique. Because of the finite values of masses of all existing charged particles and because of the smallness of the fine structure constant the pair creation process has not been observed experimentally so far.}

{In the present paper we present the derivation of this rate using both approximate semiclassical method and the method, which is based on the exact solutions of Weyl equation. In this method the field - theoretical problem of pair creation is reduced to the solution of the wave equation in the presence of a certain potential barrier. While considering this equation we encounter the signature of the so - called Klein paradox. The same problem was encountered in \cite{Nikishov}, where Nikishov considered the pair production rate for the massive Dirac fermion.
In the recent paper \cite{Gavrilov} the canonical field - theoretical formalism was developed, in which the creation operators were introduced for the specific states existing in the presence of the potential step, and the specific commutation relations were postulated (see also, \cite{Fradkin} and references therein). }
{Instead of following the methods of \cite{Nikishov,Gavrilov}, we propose our own way to resolve Klein paradox within the ranges of the pair creation problem. Our solution is based on the identification of the incoming and reflected waves opposite to the one of \cite{Nikishov}, which results from the consideration of current as a response to external electromagnetic field (the result is related somehow to the identification of waves of \cite{Ravndal}). }

\section{Relativistic fermions in Dirac/Weyl Semimetal}

\label{SectRel}

We consider the effective theory of the recently discovered Dirac semimetals $Cd_3As_2$ and $Na_3Bi$ with two Fermi (Dirac) points \cite{semimetal_discovery,semimetal_discovery2,semimetal_discovery3}.
Near each of the two Fermi points there is the pair of the left-handed and the right - handed Weyl fermions.  The action for the right - handed Weyl fermion has the form \cite{Z2015}:
\begin{equation}
    S_R = \frac{1}{2} \int d^4 \revision{x} |{\bf e}| [\bar{\Psi} i  e_b^{\revision{\mu}} {\sigma}^b {\cal D}_{\revision{\mu}}  \Psi - [{\cal D}_{\revision{\mu}}\bar{\Psi}] i  e_b^{\revision{\mu}} {\sigma}^b  \Psi ]
\label{SHe_3sW2}
\end{equation}
Here
\begin{equation}
    i{\cal D}_\mu = i\partial_\mu +  A_\mu(\revision{x})
\end{equation}
is the covariant derivative. It corresponds to the $U(1)$ gauge field $ A_\mu$.
The left - handed fermion has the action
\begin{equation}
    S_L  =  \frac{1}{2} \int d^4\revision{x} |{\bf e}| [\bar{\Psi} i  e_b^{\revision{\mu}} \bar{\sigma}^b {\cal D}_{\revision{\mu}}  \Psi - [{\cal D}_{\revision{\mu}}\bar{\Psi}] i  e_b^{\revision{\mu}} \bar{\sigma}^b  \Psi ]
\label{SHe_3sW6}
\end{equation}
Here $\bar{\sigma}^0 = 1$,  $\bar{\sigma}^a = - \sigma^a$ for $a=1,2,3$ while
\begin{equation}
    i{\cal D}_\mu = i\partial_\mu -  A_\mu(\revision{x})
\end{equation}
The vierbein in the absence of elastic deformations is given by
\begin{equation}
[e^{0}_0]^{-1}=| {\bf e}^{}|=v_F, \quad  e^{i}_a =  \hat{f}^i_a,\quad  e^{i}_0 =0, \quad e^{0}_a =0
 \label{Connection00}
\end{equation}
where $v_F$ is the average Fermi velocity, $a,i,j,k=1,2,3$, while $f^i_a=v_F \hat{f}^i_a$ has the meaning of the anisotropic Fermi velocity with the $3\times 3$ matrix
\begin{equation}
\hat{f} = {\rm diag}(\nu^{-1/3},\nu^{-1/3},\nu^{2/3}) = {\rm diag}(\hat{f}_1,\hat{f}_2,\hat{f}_3)
\end{equation}
Here we take into account that the Fermi velocity has almost the same values along the axes $X$ and $Y$. We, therefore, denote by $\nu$ the degree of anisotropy.
In practise in $Cd_3As_2$  \cite{semimetal_discovery2}  $v_F \hat{f}_1 \sim v_F \hat{f}_2 \sim c/200$ while $\hat{f}_3 \sim 0.1 \hat{f}_1$.  In   $Na_3Bi$ \cite{semimetal_discovery} $v_F \hat{f}_1 \approx 4.17\times 10^5 m/s$, $v_F\hat{f}_2 \approx 3.63 \times 10^5 m/s \sim c/800$,  while $v_F \hat{f}_3 \approx 1.1 \times 10^5 m/s$. Thus here $\hat{f}_3 \approx  0.27 \hat{f}_1$.

\section{Landau levels and the corresponding wavefunctions}

\label{SecLan}

{Let us consider massless fermions in a uniform magnetic field $\vec{H}$ directed along the third axis.
Dirac equation for the fermion with energy $\cE$ is given by
\begin{equation}
    \cE\Psi = (v_F\vec\alpha\,\hat{f}\,(-i\nabla - \vec{A})+A_0)\Psi
    \label{DEq0}
\end{equation}
where $\vec\alpha$ is in chiral representation, while $A = (A_0,\vec{A})$ is the gauge field}:
\begin{gather*}
\vec\alpha = \begin{pmatrix} \vec\sigma & 0 \\ 0 & -\vec\sigma \end{pmatrix}, \quad
\vec{A} = \begin{pmatrix} 0 \\ H y_1 \\ 0 \end{pmatrix}, \quad
    A_0 = 0
\end{gather*}
{In order to simplify the further expressions let us rescale coordinates and components of the gauge potential. We denote $y^i = v_F \hat{f}_i x^i$ ($i = 1,2,3$, sum over $i$ is not implied), and $H = \frac{1}{v^2_F \hat{f}_1 \hat{f}_2} B$.}
The {gauge} potential does not depend on coordinates transverse to $y_1$.
Hence, the solution of Dirac equation is given by the plane wave
\begin{equation}
\Psi = e^{i(p_2y_2 + p_3y_3 - \cE y_0)}\begin{pmatrix} \xi(y_1) \\ \eta(y_1) \end{pmatrix}
\end{equation}
(Here $y_0 = x_0 = t$.) With the chosen representation Eq. \eqref{DEq0} splits into
\begin{align}
    \cE\xi &=  \sigma_1(-i\xi')+\sigma_2(p_2-By_1)\xi+\sigma_3p_3\xi \label{DEq1} \\
    \cE\eta &= -\sigma_1(-i\eta')-\sigma_2(p_2-By_1)\eta-\sigma_3p_3\eta \label{DEq2}
\end{align}
{If $\cE$ is replaced by $-\cE$, the equation for $\xi$ transforms into the equation for $\eta$, and vice versa}.
For \revision{the components} $\xi_{1,2}$ we have equations similar to the one for the harmonic oscillator
\begin{align}
    \begin{split}
        (\cE^2-p_3^2+B)\xi_1 &= -\xi_1'' + B^2\left(y_1-\frac{p_2}{B}\right)^2\xi_1 \\
        (\cE^2-p_3^2-B)\xi_2 &= -\xi_2'' + B^2\left(y_1-\frac{p_2}{B}\right)^2\xi_2
    \end{split}
\end{align}
{The solutions of the system are given by $\xi_1=\psi_n(y_1-p_2/B)$ and $\xi_2=a_n\psi_{n-1}(y_1-p_2/B)$.
The corresponding energy levels are $\cE_{np_3} = \pm\sqrt{2nB+p_3^2},\enskip n \ge 0$.
$\psi_n$ are the oscillator wave functions (we assume $\psi_{-1}=0$; $H_n$ are Hermite polynomials)
\begin{equation}
    \psi_n(y) = \left(\frac{B}{\pi}\right)^{1/4}\frac{1}{\sqrt{2^nn!}}\exp\left(-\frac{B}{2}y^2\right)H_n(y\sqrt{B})
\end{equation}
The values of $a_n$ may be calculated via the direct substitution of those solutions to \eqref{DEq1}. We use the properties of Hermite polynomials and  arrive at the following system of equations
\begin{gather}
    \cE-p_3 = i a_n\sqrt{2nB}   \\
    (\cE+p_3)a_n = -i\sqrt{2nB}
\end{gather}
which is satisfied for $\cE=\cE_{np_3}$.}
The solutions of \eqref{DEq1} normalized to unity are
\begin{eqnarray}
    \xi_{np_3}(y_1) &=& \frac{1}{\sqrt{1+\frac{(\cE_{np_3}-p_3)^2}{2nB}}}\begin{pmatrix} \psi_n\left(y_1-\frac{p_2}{B}\right) \nonumber\\
      -i\frac{\cE_{np_3}-p_3}{\sqrt{2nB}}\psi_{n-1}\left(y_1-\frac{p_2}{B}\right) \end{pmatrix}\\
      && \cE_{np_3} = \pm\sqrt{2nB+p_3^2}, \quad n \ge 1
\end{eqnarray}
\begin{equation}
    \xi_{0p_3}(y_1) = \begin{pmatrix} \psi_0\left(y_1-\frac{p_2}{B}\right) \\ 0 \end{pmatrix},
      \quad \cE_{0p_3} = p_3
\end{equation}

\section{Semiclassical consideration of Schwinger pair creation in the presence of external magnetic field}

\label{SectSemiClass}

For $n\ne 0$ the pair creation process can be considered as the tunneling problem in an electric field of limited spatial extent $L$. In the present section we follow the approach of \cite{SPS}. However, our consideration differs from that of \cite{SPS}. \revision{It appears that the vacuum decay rate (unlike the pair production rate) is divergent in the massless limit.  This prompts, that the proper investigation of the massless case may require the direct consideration of massless Dirac equation rather than taking the formal limit of the expressions obtained in the massive case.}   Besides, the consideration is complicated somehow due to the anisotropy.

We add  to the system considered in section \ref{SecLan} the electric field ${E} = \frac{1}{v_F\hat{f}_3}{\tilde{E}}$ directed along the third axis in the region \revision{$0<y_3<\tilde{L}=v_F \hat{f}_3 L$ (that is $0<x_3< = L$)}.
In the time-independent gauge the component $A_0$ takes the form
\begin{equation}
    A_0 =
    \begin{cases}
        0 & \text{for } y_3 \le 0 \\
        -\tilde{E} y_3 & \text{for } 0 < y_3 <\revision{\tilde{L}} \\
        -EL & \text{for } \revision{\tilde{L}} \le y_3
    \end{cases}
\end{equation}
We imply that both $H$ and $E$ are not negative. The consideration of the opposite case is similar.

Motion of the particle that belongs to the Dirac sea ($\cE=\cE_{np_3}<0$) with $p_3>0$ and $n\ge1$ starts at $y_3<0$. $L$ is assumed to be large enough, so that $\cE+EL>0$. Weyl equation for the left-handed particles receives the form
\begin{equation}
    (\cE-A_0)\eta = i\sigma_1\partial_1\eta-\sigma_2(p_2-By_1)\eta+i\sigma_3\partial_3\eta
\end{equation}
Let us try to find its solution in the form
\begin{equation}
    \eta = \begin{pmatrix} \psi_n\left(y_1-\frac{p_2}{{B}}\right)f_1(y_3) \\
    \psi_{n-1}\left(y_1-\frac{p_2}{B}\right)f_2(y_3) \end{pmatrix}
\end{equation}
Using properties of the oscillator wave functions mentioned above we come to
\begin{equation}
    (\cE-A_0)f = \sqrt{2nB}\sigma_2f+i\sigma_3\partial_3f
\end{equation}
Instead of $y_3$ we introduce the dimensionless length $z$
\begin{gather}
    z = (\cE+\tilde{E}y_3)\sqrt{\frac{2}{\tilde{E}}}    \\
    \quad z_L=\cE\sqrt{\frac{2}{\tilde{E}}},
    \quad z_R=(\cE+{E}L)\sqrt{\frac{2}{\tilde{E}}}    \\
    \hat{z}=
    \begin{cases}
        z_L & \text{for } z \le z_L \\
        z & \text{for } z_L < z < z_R \\
        z_R & \text{for } z \ge z_R
    \end{cases}
\end{gather}
and $a=\sqrt{n\frac{B}{\tilde{E}}}$.
Next, we have
\begin{equation}
    (\sigma_1\partial_z+\frac{\hat{z}}{2}\sigma_2-a)f=0 \label{WeylEq}
\end{equation}
The Schrodinger-type equation follows
\begin{equation}
    (\partial_z^2+\frac{\hat{z}^2}{4}-a^2+\frac{i}{2}\sigma_3\partial_z\hat{z})f=0  \label{WeberEq}
\end{equation}

In order to use the semiclassical methods the turning points $\pm z_0 = \pm2a$ should be sufficiently distant, which means that $a \gg 1$. As a result we neglect the term $\frac{i}{2}\sigma_3\partial_z\hat{z}$, and obtain the semiclassical transition probability \revision{$D_{np_3} = \Big|\frac{f(z_0)}{f(-z_0)}\Big|^2$:}
\begin{equation}
    D_{np_3} = \exp\left(-2\int_{-z_0}^{z_0}dz\sqrt{a^2-z^2/4}\right) = \exp\left(-2\pi a^2\right)
\end{equation}
It may be interpreted as the exponential factor in the probability that the pair particle - hole is created at the $n$ - th level.

In order to have the incoming wave that corresponds to the negative energy we need
$z_L< -2a$, while in order to have the outgoing wave with positive energy we need $z_R>2 a$. Together these requirements give
\begin{equation}
{E}L - \sqrt{2 n B} >|\cE | > \sqrt{2 n B}\label{cond}
\end{equation}
Hence, the values of $p_3$ should satisfy $0 \le p_3 \le p^{(c)}$, $p^{(c)} \approx EL$.
\revision{To obtain the overall semiclassical probability for the creation of pairs, we integrate over the \(n\)-th Landau level with the density of states \(\frac{HL_1L_2|}{4\pi^2}\), where \(L_1, L_2\) are the corresponding sizes of the sample, and the sum is over all levels}
\begin{equation}
    N = \tilde{L} \sum_n\int^{+p^{(c)}}_{0} \frac{dp_3}{2\pi} \frac{HL_1L_2}{2\pi}\exp\left(-2\pi n\frac{B}{\tilde{E}}\right)
\end{equation}
Here we substituted $\tilde{L} = T$, where by $T$ we denote the overall duration of the process.

Recall, that the same consideration may be applied to the right-handed particles.
Finally we come to the expression for the pair production rate in the unit volume at the $n$ - th Landau level:
\begin{equation}
    \dot{\rho}_n = \frac{dN}{Vdt} =   \frac{  EH}{2\pi^2}\exp\left(-2\pi n v_F \nu^{-4/3}\frac{H}{E}\right)\label{rhon}
\end{equation}
{In order to rewrite this expression in covariant form let us introduce the notations:
\begin{eqnarray}
    {\cal S} &=& \frac{1}{4} F_{ij}F_{kl}g^{ik}g^{jl}\nonumber\\
    {\cal P} &=& \frac{1}{\sqrt{-g}}\frac{1}{8} F_{ij}F_{kl}\epsilon^{\revision{ijkl}}\label{SP}
\end{eqnarray}
where
$$g^{ij} = e^i_ae^j_b\eta_{ab} = \left(\begin{array}{cccc}1/v^2_F&0&0&0\\0&-\nu^{-2/3}&0&0\\0&0&-\nu^{-2/3} \\0&0&0&-\nu^{4/3}\end{array} \right)$$ and $\eta_{ab}$ is metric of Minkowski space while $g$ is the inverse determinant of matrix $g^{ij}$. Notice, that $\sqrt{-g} = |{\bf e}| = v_F$. Using these notations we rewrite Eq. (\ref{rhon}) in the form
\begin{equation}
    \dot{\rho}_n = \frac{dN}{Vdt} =   \sqrt{-g}\frac{|{\cal P}|}{2\pi^2}\exp\left(-2\pi n \frac{|{\cal P}|}{\sqrt{{\cal S}^2+{\cal P}^2}-{\cal S}}\right)
\end{equation}}

\section{Exact consideration of Schwinger pair creation in the presence of external magnetic field}

\label{SectExact}

The two independent solutions of Weber's equation \eqref{WeberEq} in the region $z_L<z<z_R$ may be written in the form (for the second component of $f$):
\begin{gather}
    M(a,z)=\exp\left(i\frac{z^2}{4}\right)\Phi\left(\frac{a^2i}{2},\frac{1}{2},-i\frac{z^2}{2}\right) \\
    N(a,z)=z\exp\left(i\frac{z^2}{4}\right)\Phi\left(\frac{1+a^2i}{2},\frac{3}{2},-i\frac{z^2}{2}\right)
\end{gather}
where $\Phi$ is the confluent hypergeometric function.
For the first component of $f$ the two independent solutions are $M^*$ and $N^*$.
We can show that
\begin{align}
    \left(\partial_z-i\frac{z}{2}\right)M &= a^2 N^*\\
    \left(\partial_z-i\frac{z}{2}\right)N &= M^*
\end{align}
And the general solution of Eq. \eqref{WeylEq} is given by
\begin{equation}
    f = \begin{pmatrix} AM^*+aBN^* \\ BM+aAN \end{pmatrix}
\end{equation}
At $z<z_L$ and $z>z_R$ the solutions of Eq. \eqref{WeylEq} are the plane waves propagating in opposite directions
\begin{eqnarray}
    g&=&A_{L,R} \chi_+ e^{ikz}+B_{L,R} \chi_- e^{-ikz}\nonumber\\
    k&=&k_{L,R} = \sqrt{\frac{z_{L,R}^2}{4}-a^2}\nonumber\\
    \chi_\pm &=& \frac{1}{N_{L,R}^\pm}\begin{pmatrix}i(\pm k-\frac{z_{L,R}}{2})\\a\end{pmatrix}
\end{eqnarray}
with $N^\pm_{L,R}=\sqrt{a^2 +(\pm k-\frac{z_{L,R}}{2})^2} $

In the region $z<z_L$ there are both the incoming wave and the reflected wave, while for $z>z_R$ there is only the outgoing one, which means that
\begin{align}
    e^{ik z}\chi_+ + Re^{-ik z}\chi_- &= f \qquad \text{at } z=z_L, k=k_L\\
        Te^{ik z}\chi_+ &= f \qquad \text{at } z=z_R , k=k_R
\end{align}
The phase factors may be absorbed by the redefinition of $T$ and $R$, which gives
\begin{eqnarray}
\left(\begin{array}{cc}M^*_R&a N^*_R\\aN_R & M_R\end{array}\right)\left(\begin{array}{c}A\\B\end{array}\right) & = & \frac{T}{N^+_R} \left(\begin{array}{c}i(k_R - z_R/2)\\a \end{array}\right)\label{AB}\\
\left(\begin{array}{cc}M^*_L&a N^*_L\\aN_L & M_L\end{array}\right)\left(\begin{array}{c}A\\B\end{array}\right) &=& \left(\begin{array}{c}\frac{1}{N^+_L} i (k_L - \frac{z_L}{2})-\frac{R}{N^-_L}(k_L + \frac{z_L}{2})\\a (1/N_L^++R/N_L^-) \end{array}\right)\nonumber
\end{eqnarray}
We have $z_R \gg 1$.
Let us also suppose, that $|z_L| \gg 1$. (The region with $|z_L|\gg 1$ dominates in the integral over momenta for the pair production rate if  the system size is sufficiently large, such that $L \sqrt{\frac{E}{v_F\hat{f}_3}} \gg 1$.) Then we have the asymptotic expressions:
\begin{eqnarray}
M_{R,L} & \approx & \frac{\sqrt{\pi}}{\Gamma((1-ia^2)/2)}\,e^{\frac{\pi}{4}a^2}\, e^{i(z^2/4-
\frac{a^2}{2}{\rm log}z^2/2)}\nonumber\\
N_{R,L} & \approx & \frac{\mp i\sqrt{\pi}}{\sqrt{2}\Gamma((2-ia^2)/2)}\,e^{\frac{\pi}{4}a^2}\, e^{i(z^2/4-
\frac{a^2}{2}{\rm log}z^2/2)}
\end{eqnarray}
Under the same conditions Eq. (\ref{AB}) is simplified.
We derive
\begin{equation}
    \abs{T}^2= \frac{1}{\exp\left(2\pi n v_F \nu^{-4/3}\frac{H}{E}\right) - 1}
\end{equation}
and
\begin{equation}
    \abs{R}^2=\frac{\exp\left(2\pi n v_F \nu^{-4/3}\frac{H}{E}\right)}{\exp\left(2\pi n v_F \nu^{-4/3}\frac{H}{E}\right) - 1}
\end{equation}
{One can see, that $$ \abs{R}^2 = 1 +  \abs{T}^2>1$$ which is related to Klein paradox and does not allow to interpret $ \abs{R}^2$ as the reflection probability. The same refers also to  $\abs{T}^2$, which may become larger, than unity at sufficiently small $H$. The possible removal of this paradox was explained in \cite{Nikishov1970} (for the modern development of this approach see \cite{Gavrilov}). Nikishov \cite{Nikishov1970} proposed to consider $\abs{T}^2$ as the so - called relative pair production probability. In order to calculate the absolute pair production probability he proposed to multiply this value by the quantity $\abs{C}^2$, which is related to $ \abs{T}^2$ as follows: \begin{equation}
\abs{C}^2 +  \abs{C}^2\times  \abs{T}^2= 1 \label{nikishov}
\end{equation}
The left hand side of this equation was interpreted as the sum of the probability that no pairs are created $ \abs{C}^2 $ and the probability $ D_{np_3}= \abs{C}^2\times  \abs{T}^2$ that one pair is created in the given state. The origin of this factor was not described in \cite{Nikishov}. In \cite{Gavrilov} the modification of the canonical quantization for the field theory in the presence of potential steps was proposed, in which such factor  $ \abs{C}^2 $ appears in a natural way. To the best of our knowledge no rigorous derivation of this heuristic rule exists. However, it allows to reproduce the result of Schwinger \cite{Schwinger} for the vacuum decay probability of the massive Dirac field obtained originally in the essentially different formalism.}

{Here we propose alternative and very simple explanation of the Nikishov's prescription. Let us restrict ourselves by consideration of the left - handed fermions. (The consideration of the right - handed ones is similar.) We look at the electric current density (equal to the spin current density with the minus sign) at both sides of the potential barrier. Right to the barrier, where the wave with the amplitude $ \abs{T}^2$ appears, the electric current of the left - handed particles is given by
\begin{equation}
j_T = -f^+ \sigma^3 f =  \abs{T}^2
\end{equation}
while left to the barrier, there are the incoming and the reflected waves. If those two waves are considered separately, the corresponding currents are given by
\begin{equation}
j_R = -f^+ \sigma^3 f =  \abs{R}^2, \quad j_i =- f^+ \sigma^3 f = - 1
\end{equation}
Therefore, the interpretation of the components of the spinor left to the barrier is inverse to the naive one. Although the wave vector for the first component of the spinor is positive, the corresponding current is directed to the opposite direction. That's why, this is the reflected rather than the incoming wave. (This corresponds to the change $p \rightarrow -p$ in the antiparticle region proposed in \cite{Ravndal}.) Thus, we should normalize our solution in such a way, that the amplitude of the second component of the spinor left to the barrier is equal to unity. This gives immediately the transition coefficient
\begin{equation}
    D_{np_3}= \frac{\abs{T}^2}{1+\abs{T}^2} = \exp\left(-2\pi n v_F \nu^{-4/3}\frac{H}{E}\right)
\end{equation}
which coincides with the semiclassical expression obtained in the previous section. Notice, that the same expression may be obtained if the formalism of \cite{Gavrilov} would be applied to the consideration of the given problem. Recall, that this is the Hamiltonian formalism, which is based on the specific choice of the in and out states in the presence of the potential step and on the postulation of the corresponding commutation relations.}

{The total pair production rate \(\Gamma\) includes the contribution of anomaly that originates from the lowest Landau level:
\begin{eqnarray}
    \Gamma &=& \frac{2}{VT}\sum_{n=1}^{\infty}\int\frac{dp_3}{2\pi}T\frac{H L_1L_2}{2\pi}D_{np_3}+\frac{EH}{4\pi^2} \nonumber\\
       & = &  \frac{EH}{4\pi^2} {\rm cth}\left(\pi  v_F \nu^{-4/3}\frac{H}{E}\right) \label{Gamma}
\end{eqnarray}}
\revision{One can see, that as well as Eq. (3) of \cite{SPS} our Eq. (\ref{Gamma}) gives  the pair production rate that is not vanishing in the limit \(H\to 0\). Thus, our results also confirm the corresponding expression obtained in the case of the vanishing magnetic field \cite{SPS}.}

\section{Pair production rate and vacuum persistence probability}

\label{SectPers}

Vacuum persistence probability of any 4-volume \(\Omega=\Delta V\Delta t\) under the conditions described above is given by \cite{Schwinger}, \cite{Nikishov}, \cite{SPS}
\begin{equation}
    P=e^{-w\Omega}
\end{equation}
where {
\begin{equation}
    w = \frac{EH}{4\pi^2} \sum_{k = 1}^{\infty} \frac{1}{k} {\rm cth} \left(k\,v_F \nu^{-4/3}\frac{H}{E} \right) \exp\left(- \frac{k \pi m^2}{v_F \nu^{2/3}E} \right) \label{w}
\end{equation}}
and \(m\) is the Dirac mass. One can see, that \(w\) diverges in the limit \(m \to 0\), which makes the vacuum persistence probability equal to zero. Of course, this is not a surprise since the spectral flow of the one particle Hamiltonian provides the definite appearance of the pairs of the left - handed and the right - handed particles on the lowest Landau level. As a result the vacuum definitely cannot survive.

{Nevertheless, this result does not imply necessarily, that the pair production rate is infinite. Indeed we have seen in the previous section, that it is given by the finite expression of Eq. (\ref{Gamma}) that gives the first term in the sum over $k$ in Eq. (\ref{w}).} {According to Nikishov \cite{Nikishov} the vacuum decay rate should be calculated as
\begin{eqnarray}
 2 \, {\rm Im} \, {\cal L}  \equiv w &=& \frac{2}{VT}\sum_{n=1}^{\infty}\int\frac{dp_3}{2\pi}T\frac{HL_1L_2}{2\pi}{\rm log}(1-D_{np_3}) \nonumber\\&&+ \frac{1}{VT}\int\frac{dp_3}{2\pi}T\frac{HL_1L_2}{2\pi}{\rm log}(1-D_{ 0 p_3})
\end{eqnarray}
For the massless case $D_{ 0 p_3} = 1$, which means, that on the lowest Landau level the pair is created with the probability equal to unity. As a result the imaginary part of the effective lagrangian is formally divergent.}

\section{Contribution to conductivity due to the pair creation (the ideal non - interacting system at $T=0$)}

\label{SectCond}

{Particle is created with energy $\cE^\prime = \cE + LE$ (counted from the level $-EL$) and momentum $p^\prime_3 = \sqrt{(\cE + LE)^2 - 2 n B }$. In order to calculate the values of momentum $p_3^{\prime\prime}$ carried by the hole, let us use equation $\frac{d}{dt}\langle p_3 \rangle = \tilde{E}^3$. If the tunneling does not occur, then the evolution in time of the given occupied state is governed by this equation. If we start from the state with momentum $p_3$, its average over the given state is changed with time. We take into account that the overall duration of the process $T$ is equal to length $L$ divided by the Fermi velocity $v_F \nu^{2/3}$, which is the velocity of the given particle:  $T = \tilde{L}$.  Therefore,
\begin{equation}
\langle p_3^{\prime\prime}\rangle = p_3+\tilde{E} \tilde{L}
\end{equation}
As a result, the positive energy of the hole (created together with the electron) may be estimated as:
\begin{equation}
\langle \cE^{\prime\prime}\rangle \approx \sqrt{(EL)^2+\cE^2 + 2 EL \sqrt{\cE^2-2nB}} \approx EL + |\cE|
\end{equation}}
The Schwinger pair creation costs energy. Suppose, that the single pair of particle with energy $\cE + LE$ and hole with energy $\approx |\cE| + EL$ is created. This requires energy $2LE$. {In the ideal system the created pairs do not annihilate. Therefore, their number and the filling factors of the excited states are increased with time. In general case, the Pauli principle affects the whole process\footnote{Each state cannot be filled by more than one fermion. At the later stages of the pair creation the filling factors of the excited states (holes at ${\cal E}<0$ and particles at ${\cal E}>0$) become finite, and we have to take into account Pauli interdiction while considering the pair production process and the corresponding energy balance.}. Let us assume, that we are at the beginning of the pair creation process, when the filling factors of the excited states are still small. Then we may neglect the Pauli interdiction. } The total energy needed to produce $$\Delta N =  \frac{  { E}{H}}{4\pi^2}\Big(1+2\sum_{n=1}^\infty\exp\left(-2\pi v_F \nu^{-4/3}\frac{H}{E}\right) \Big) L_1L_2L_3\Delta t$$ pairs is
equal to ${ j}^{(1)}{E}V\Delta t$, where ${ j}^{(1)}$ is the contribution to the total electric current related to the pair production process. As a result, we have
\begin{equation}
{j}^{(1)} =2 L \frac{ {E}{ H}}{4\pi^2}\, {\rm cth} \left(\pi \,v_F \nu^{-4/3}\frac{H}{E} \right) \label{je}
\end{equation}
(Here we considered the contribution of one Dirac fermion.) Of course, this expression is to be modified in the real systems.

\section{A rough consideration of particle - hole plasma}
\label{SectPlasma}

{In practise the quasiparticles interact with each other. Besides, there are the finite temperature fluctuations. During the competition of the Schwinger pair creation and the annihilation of pairs during the collisions of quasiparticles the particle - hole plasma is formed.}

{In \cite{ZrTe5} only the pairs of the left - handed and the right - handed quasiparticles were considered. Their accumulation was described by the
kinetic equation
\beqn
\frac{d \rho_0}{d t}  = - \frac{\rho_0}{\tau_0}   + \frac{1}{4 \pi^2} HE
\label{eq:drho5}
\eeqn
where $\rho_0$ is the density of pairs (left - handed particle plus right - handed antiparticle) incident at the lowest Landau level. The first term in the right hand side describes the impact of the collisions with change of chirality. Those collisions may occur between the fermionic quasiparticles, or between the fermionic quasiparticles and impurities, or between the considered fermionic quasiparticles and the other excitations existing in the semimetals. The corresponding average scattering time is denoted by $\tau_0$. The second term describes generation of the chiral charge due to the chiral anomaly.  The solution of Eq. (\ref{eq:drho5}) at $t\gg \tau_0$ gives
\begin{equation}
\rho_0 = \frac{\tau_0}{4 \pi^2} HE\label{TAU0}
\end{equation}}

{The lowest Landau level dominates if $v_F H \gg E$. The correction to Eq. (\ref{eq:drho5}) due to the pair production at the higher Landau levels may be estimated as well. As a result the system of equations for different densities should appear. However, we do not consider here this complication and will concentrate on the corresponding contribution to conductivity. We introduce the extra relaxation time scale $\tau_{1}\ll \tau_0$ that corresponds both to the processes with annihilation of the particles and the anti - particles of the same chiralities, and to the scattering without change of chirality of the fermionic quasiparticles on each other, on impurities, and on the other excitations existing in the semimetals. In order to be able to apply the considered above formalism to the calculation of the pair production rate we need \begin{equation}
E  \tau_1 \gg \sqrt{2  H}\label{TAU00}
\end{equation}
 In this case instead of the length of the sample entering Eq. (\ref{je}) we substitute the mean free path\footnote{This mean free path is taken along the common direction of electric and magnetic fields, which are directed here along the third axis.}, that is given by $v_F \nu^{2/3}\tau_1$.  In general case the scattering time $\tau_1$ may depend on the densities of the quasiparticles and thus on $E$ and $H$. For simplicity we neglect here such a dependence. 
This gives the following correction to conductivity
\begin{eqnarray}
\sigma^{(1)} &\approx & N_D v_F \nu^{2/3} \tau_1 \frac{ H}{2\pi^2}\times \Big[{\rm cth} \left(\pi\,v_F \nu^{-4/3}\frac{H}{E} \right) \nonumber\\&&- \pi  v_F \nu^{-4/3}\frac{H}{E}\Big(1- {\rm cth}^2 \left(\pi\,v_F \nu^{-4/3}\frac{H}{E} \right) \Big)\Big]
\label{sigma2}
\end{eqnarray}
Here $N_D$ is the number of Dirac points of the given semimetal. In particular, for Cd$_3$As$_2$ and Na$_3$Bi $N_D=2$. At $v_F H \gg E$ this expression gives the contribution of chiral anomaly,
which differs essentially from the expression obtained in \cite{ZrTe5}.

 {In  \cite{ZrTe5} it was argued, that $\sigma^{(1)} \sim H^2$ (for the discussion of this issue see also \cite{Chernodub:2015wxa,semimetal_effects6,ref:diffusion,ref:transport,ZrTe5:2,ChiralAnomalySemimetal}). \cite{ZrTe5} appeals to the semi - equilibrium pattern, in which the chiral chemical potential $\mu_5$ is generated that is related to $\rho_0$ for $T \gg \mu_5$ as follows: $\mu_5 = \frac{6\,v_F^3 }{T^2}\rho_0$. Next, the
 expression for the chiral magnetic effect \cite{CME}  $j = \frac{\mu_5}{2\pi^2}H$ was applied, which gives $\sigma^{(1)} \sim H^2$. However, it was demonstrated recently that the equilibrium chiral magnetic effect in the real semimetals does not exist \cite{noCME}.}

{Here we propose the alternative model, which is based on the simple energy balance:  the energy needed to produce the pairs should be taken from the work $Ej$ performed by the electric field. In order to apply this balance we should be able to neglect Pauli interdiction. To do this we need that the filling factors are small for the states, in which the pairs are created. This occurs for
\begin{equation}
T \ll v_F E \tau_1 \label{TAU1}
\end{equation}
 when the Fermi distribution for those states gives $\frac{1}{e^{v_F E \nu^{2/3}\tau_1/T}+1}\ll 1$.  Being created with the energy $2 v_F E \nu^{2/3} \tau_1$ the pairs are thermalized and those states become almost empty again.  Thus the energy balance gives Eqs. (\ref{je}) and (\ref{sigma2}). In the opposite limit, when $T \gg v_F E \tau_1$ the Fermi distribution gives the filling factor $\approx 1/2$. The creation of particles/holes is not prohibited completely in these states. However, Pauli principle reduces the pair production rate. Then the contribution to conductivity may still be given by Eq. (\ref{sigma2}) with the reduction factor.}

{Finally, let us recall, that in addition to $\sigma^{(1)}$ there exists the ordinary Ohmic contribution to conductivity $\sigma^{(0)}$, which is not related directly to the pair production process, and which is caused by  existence of the thermal charge carriers.}

\section{Discussion}

\label{SectCon}
{In the present paper we calculated the pair production rate using the solution of the corresponding quantum mechanical problem. Unlike the approach of \cite{Gavrilov} we do not attempt to construct the specific axiomatic quantum field theory with unstable vacuum. According to our point of view this is not needed as a matter of principle for the consideration of the dynamics of electrons in solids, which are exhaustively described by the multi - particle quantum mechanics. In the vicinity of the Fermi point in Dirac semimetals this quantum mechanics is described approximately by massless Dirac equation, that is the couple of the Weyl equations.  }

We suppose, that the exhaustive consideration of the contribution of the pair creation to conductivity will require essential efforts. This consideration should include the application of the contemporary methods of kinetic theory. Nevertheless, we expect, that our approximate consideration is relevant for the description of the system at least in the region of parameters given by Eqs. (\ref{TAU00}), (\ref{TAU1}). This approximate consideration is based on the synthesis of the solution of the simple quantum mechanical problem (with the interaction between the quasiparticles neglected) and the phenomenological consideration of the non - equilibrium pair creation process based on the application of the notion of the mean free path of the quasiparticles. The specific dependence of Eq. (\ref{sigma2}) on electric and magnetic fields may, in principle, be extracted from the dependence on these fields of the total conductivity. This may allow to observe for the first time the Schwinger pair production in the three - dimensional systems in laboratory.  (Previously the similar possibility to observe the Schwinger pair production in graphene was discussed extensively - see, for example, \cite{Schwingergraphene,Zubkov:2012ht} and references therein.) \revision{In principle, the Schwinger pair creation takes place also in the condensed matter systems with gapped fermions. For example, it is present in the Dirac insulators that may be described effectively by the low energy theory with massive Dirac fermions in a certain approximation. In this case, however, unlike the case of Dirac semimetals the pair production probability is suppressed essentially by the exponential factor containing the fermion mass just like in the high energy physics, where all known charged fermions are massive.}

\section*{Acknowledgements}
The  part of the work of  M.A.Z. performed in Russia (Sections \ref{SectRel}, \ref{SecLan}, \ref{SectSemiClass}, \ref{SectCond}) was supported by Russian Science Foundation Grant No 16-12-10059 while the part of the work made in France (Sections  \ref{SectExact}, \ref{SectPers}, \ref{SectPlasma}) was supported by Le Studium Institute of Advanced Studies.


\begin{thebibliography}{99}



\bibitem{semimetal_discovery}
Z. K. Liu et al.,
{\sl ``Discovery of a Three-dimensional Topological Dirac Semimetal, Na${}_3$Bi''},
Science (2014) {\bf 343}, 864 [arXiv:1310.0391].

\bibitem{semimetal_discovery2}
M. Neupane et al.,
{``Observation of a topological 3D Dirac semimetal phase in high-mobility Cd${}_3$As${}_2$''}
Nature Commun. {\bf 05}, 3786 (2014) [arXiv:1309.7892].

\bibitem{semimetal_discovery3}
S. Borisenko et al.,
{``Experimental Realization of a Three-Dimensional Dirac Semimetal''},
Phys. Rev. Lett. {\bf 113}, 027603 (2014) [arXiv:1309.7978].

\bibitem{ZrTe5}
Q. Li, D. E. Kharzeev, C. Zhang, Y. Huang, I. Pletikosic, A. V. Fedorov, R. D. Zhong, J. A. Schneeloch, G. D. Gu, and T. Valla,
arXiv:1412.6543.

\bibitem{ZrTe5:2}
R. Y. Chen, S. J. Zhang, J. A. Schneeloch, C. Zhang, Q. Li, G. D. Gu, N. L. Wang,
{\sl ``Optical spectroscopy study of three dimensional Dirac semimetal ZrTe$_5$''},
arXiv:1505.00307.

\bibitem{Bi2Se3}
Devendra Kumar, Archana Lakhani,
{\sl ``Observation of three-dimensional Dirac semimetal state in topological insulator Bi$_2$Se$_3$''},
arXiv:1504.08328.

\bibitem{WeylSemimetalDiscovery}
B. Q. Lv et al.,
{\sl ``Experimental discovery of Weyl semimetal TaAs''},
Phys. Rev. X 5, 031013 (2015), arXiv:1502.04684;
X. Huang,
{\sl ``Observation of the chiral anomaly induced negative magneto-resistance in 3D Weyl semi-metal TaAs''}, Phys. Rev. X 5, 031023 (2015), arXiv:1503.01304;
B. Q. Lv et al.,
{\sl ``Observation of Weyl nodes in TaAs''},
arXiv:1503.09188.

\bibitem{Volovik2003}
G.E. Volovik,
{\sl ``The Universe in a Helium Droplet''},
Clarendon Press, Oxford (2003)

\bibitem{semimetal_effects10}
M. Vazifeh and M. Franz,
{\sl ``Electromagnetic response of weyl semimetals''},
Phys. Rev. Lett. {\bf 111}, 027201 (2013) [arXiv:1303.5784].

\bibitem{semimetal_effects11}
Y. Chen, S. Wu, and A. Burkov,
{\sl ``Axion response in Weyl semimetals''},
Phys. Rev. B {\bf 88}, 125105 (2013) [arXiv:1306.5344].

\bibitem{semimetal_effects12}
Y. Chen, D. Bergman, and A. Burkov,
{\sl ``Weyl fermions and the anomalous Hall effect in metallic ferromagnets''},
Phys. Rev. B {\bf 88}, 125110 (2013) [arXiv:1305.0183];
David Vanderbilt, Ivo Souza, and F. D. M. Haldane
Phys. Rev. B {\bf 89}, 117101 (2014) [arXiv:1312.4200].

\bibitem{semimetal_effects13}
S. T. Ramamurthy and T. L. Hughes,
{\sl ``Patterns of electro-magnetic response in topological semi-metals''},
arXiv:1405.7377.

\bibitem{ref:semimetal:4}
L. P. He et al.,
{\sl Quantum Transport Evidence for the Three-Dimensional Dirac Semimetal Phase in Cd${}_3$As${}_2$},
Phys. Rev. Lett. {\bf 113}, 246402 (2014) [arXiv:1404.2557].



\bibitem{Zyuzin:2012tv}
A.~A.~Zyuzin and A.~A.~Burkov,
  {\sl ``Topological response in Weyl semimetals and the chiral anomaly,''}  Phys.\ Rev.\ B {\bf 86} (2012) 115133  [arXiv:1206.1868 [cond-mat.mes-hall]].  


\bibitem{tewary}
Pallab Goswami, Sumanta Tewari, {\sl Axionic field theory of (3+1)-dimensional Weyl semi-metals,}
Phys. Rev. B 88, 245107 (2013), arXiv:1210.6352

\bibitem{Chernodub:2015wxa}
  M.N.Chernodub and M.Zubkov,
{\sl ``Intrinsic chiral magnetic effect in Dirac semimetals due to dislocations,''}
  arXiv:1508.03114 [cond-mat.mes-hall].

\bibitem{ref:diffusion}
Cheng Zhang, Enze Zhang, Yanwen Liu, Zhi-Gang Chen, Sihang Liang, Junzhi Cao, Xiang Yuan, Lei Tang, Qian Li, Teng Gu, Yizheng Wu, Jin Zou, Faxian Xiu,
{\sl ``Detection of chiral anomaly and valley transport in Dirac semimetals''},
arXiv:1504.07698.

\bibitem{ref:transport}
Tian Liang,Quinn Gibson, Mazhar N. Ali, Minhao Liu, R. J. Cava, N. P. Ong,
{\sl Ultrahigh mobility and giant magnetoresistance in the Dirac semimetal Cd${}_3$As${}_2$},
Nature Mater. {\bf 14}, 280 (2015) [arXiv:1404.7794].

\bibitem{SMR}
Thomas D. Cohen, David A. McGady
{\sl ``The Schwinger mechanism revisited'' },
Phys.Rev.D78:036008,2008, \href{http://arxiv.org/abs/0807.1117}{arXiv:0807.1117}.

\bibitem{Schwinger}  J.~Schwinger,  Phys.\ Rev.\  {\bf 82}, 664 (1951).

\bibitem{Sauter} F. Sauter, Z. Phys. 69, 742 (1931); W. Heisenberg and H. Euler, Z. Physik 98, 714 (1936); V. Weisskopf, K. Dan. Vidensk.
Selsk. Mat. Fys. Medd. 14, No. 6 (1936).

\bibitem{Nikishov} A.I. Nikishov {\sl ``Pair Production by a Constant External Field''}, \href{http://www.jetp.ac.ru/cgi-bin/dn/e_030_04_0660.pdf}{JETP, Vol. 30, No 4, p. 660 (April 1970)}.

\bibitem{SPS} S.~P.~Kim and D.~N.~Page,
  ``Schwinger pair production in electric and magnetic fields,'' Phys.Rev.D73:065020,2006, \href{http://arxiv.org/abs/hep-th/0301132}{arxiv:hep-th/0301132v3}.



\bibitem{Nikishov1970}
A.I. Nikishov, "Barrier scattering in field theory. Removal of Klein paradox," Nuclear Physics B21 (1970) 346-358

\bibitem{Ravndal}A. Hansen, F. Ravndal, Phys. Scripta 23 (1981) 1033

\bibitem{Gavrilov}
  S.~P.~Gavrilov and D.~M.~Gitman,
  ``Vacuum polarization and particle creation in the presence of a potential step,''
  Int.\ J.\ Mod.\ Phys.\ A {\bf 31} (2016) no.02n03,  1641031.
  doi:10.1142/S0217751X16410311

\bibitem{Fradkin}E. S. Fradkin,
D. M. Gitman, and S. M. Shvartsman, Quantum Electrodynamics
with Unstable Vacuum (Springer-Verlag, Berlin,
1991).

\bibitem{Nielsen:1983rb}
  H.~B.~Nielsen and M.~Ninomiya,
  ``Adler-bell-jackiw Anomaly And Weyl Fermions In Crystal,''
  Phys.\ Lett.\ B {\bf 130} (1983) 389.
  doi:10.1016/0370-2693(83)91529-0

\bibitem{CME}
 K. Fukushima, D.E. Kharzeev, H.J. Warringa, Phys.Rev.D 78:074033,2008

\bibitem{ChiralAnomalySemimetal}
Hui Li, Hongtao He, Hai-Zhou Lu, Huachen Zhang, Hongchao Liu, Rong Ma, Zhiyong Fan, Shun-Qing Shen, Jiannong Wang,
{\sl ``Negative Magnetoresistance in Dirac Semimetal Cd${}_3$As${}_2$''},
arXiv:1507.06470;

\bibitem{semimetal_effects6}
S. Parameswaran, T. Grover, D. Abanin, D. Pesin, and A. Vishwanath,
{\sl ``Probing the chiral anomaly with nonlocal transport in Weyl semimetals},
Phys. Rev. X {\bf 4}, 031035 (2014) [arXiv:1306.1234].

\bibitem{noCME}
Mikhail Zubkov, M.A.Zubkov,{\sl  "Absence of equilibrium chiral
magnetic effect"}, arXiv:1605.08724, Physical
Review D 93, 105036 (2016), $<$hal-01275180v3$>$,
https://hal.archives-ouvertes.fr/hal-01275180v3/document

\bibitem{Z2015}
M.A.Zubkov,
{\sl ``Emergent gravity and chiral anomaly in Dirac semimetals in the presence of dislocations''},
Annals of Phys., {\bf 360}, 655 (2015), [arXiv:1501.04998]

\bibitem{Schwingergraphene}
  M.~I.~Katsnelson, G.~E.~Volovik and M.~A.~Zubkov,
  ``Euler - Heisenberg effective action and magnetoelectric effect in multilayer graphene,''
  Annals Phys.\  {\bf 331} (2013) 160
  doi:10.1016/j.aop.2012.12.010
  [arXiv:1206.3973 [cond-mat.mes-hall]].


\bibitem{Zubkov:2012ht}
  M.~A.~Zubkov,
  ``Schwinger pair creation in multilayer graphene,''
  Pisma Zh.\ Eksp.\ Teor.\ Fiz.\  {\bf 95} (2012) 540
  doi:10.1134/S0021364012090135
  [arXiv:1204.0138 [hep-ph]].

\end{thebibliography}
\end{document}